# Sparse Oblique Decision Tree for Power System Security Rules Extraction and Embedding

Qingchun Hou, *Student Member, IEEE*, Ning Zhang, *Senior Member, IEEE*, Daniel S. Kirschen, *Fellow, IEEE*, Ershun Du, *Member, IEEE*, Yaohua Cheng, *Student Member, IEEE,* Chongqing Kang, *Fellow, IEEE*

*Abstract*—Increasing the penetration of variable generation has a substantial effect on the operational reliability of power systems. The higher level of uncertainty that stems from this variability makes it more difficult to determine whether a given operating condition will be secure or insecure. Data-driven techniques provide a promising way to identify security rules that can be embedded in economic dispatch model to keep power system operating states secure. This paper proposes using a sparse weighted oblique decision tree to learn accurate, understandable, and embeddable security rules that are linear and can be extracted as sparse matrices using a recursive algorithm. These matrices can then be easily embedded as security constraints in power system economic dispatch calculations using the Big-M method. Tests on several large datasets with high renewable energy penetration demonstrate the effectiveness of the proposed method. In particular, the sparse weighted oblique decision tree outperforms the state-of-art weighted oblique decision tree while keeping the security rules simple. When embedded in the economic dispatch, these rules significantly increase the percentage of secure states and reduce the average solution time.

## I. INTRODUCTION

The increased uncertainty on the net load pattern that results from a high penetration of variable renewable energy generation can substantially change the operating modes of a power system and make the potential failure mechanisms more diverse [1]. A larger proportion of PV and wind generation reduces the system's inertia and reduces its ability to mitigate contingencies[2][3]. In addition, extreme events are more likely to happen, such as unexpected cascading N-k contingencies [4], small signal instabilities, and voltage instabilities [5]. It is thus becoming harder to define accurate and understandable rules that can be used to reliably assess the security of the system and substantiate preventive control actions [6]. Defining these security boundaries and embedding them in the economic dispatch is thus essential to the reliable and optimal operation of power systems with a high renewable penetration. While some operational boundaries can be expressed by analytical expressions, this is not always possible, particularly when considering the stochastic nature of operation. One could instead extract rules or constraints based on simulated system operating data.

A number of authors [7][8][9][10] have used decision trees (DT) to learn rules for assessing the dynamic or voltage stability of power system operating states. Recently, researchers have developed techniques to embed these rules in power system optimization programs [11][12][13].

Most of these previous studies use univariate decision tree (UDT) such as CART [14], ID3 [15], C4.5 [16] for defining these rules. However, UDTs have several limitations that hinder extraction of accurate and embeddable rules:

1) Because UDTs split on only one feature at each node, their learning capacity is limited and more likely to result in under-fitting when learning complex rules with a shallow tree;

2) Because the knowledge UDT acquire at each node is limited, they tend to be very deep, with an exponentially increasing number of leaves. Such a complex structure makes the rules impossible to understand. They are also hard to embed in optimizations because each leaf must be modeled as an integer variable in optimization;

3) Because at each node the candidate split value is chosen from the feature values in the samples, UDT rules can alter significantly with minor changes in only a few samples.

To overcome these limitations, this paper proposes a technique to develop *sparse weighted oblique decision trees* (SWODT) to learn sparse security rules. The Orthant-Wise Limited-memory Quasi-Newton (OWL-QN) method is modified to solve the non-smooth optimization problem required to obtain the split parameters at each node. We also propose a recursive method to extract sparse matrix rules from the sparse weighted oblique decision tree. The Big-M method is then used to embed these sparse matrix rules as security constraints in a Mixed Integer Linear Programming (MILP) power system economic dispatch.

The remainder of the paper is organized as follows. Section II briefly reviews the literature on the application of decision trees to power system security assessment and security-constrained economic dispatch. Section III describes the data-driven framework used in this paper. Section IV introduces the SWODT and its training algorithm. Section V explains how to extract sparse matrix rules and embed them in economic dispatch. Section VI demonstrates the effectiveness of the proposed method on several test systems with a high penetration of renewable generation. Section VII concludes.

This work is supported in part by the National Key Research and Development Program of China (No. 2016YFB0900100). Q. Hou, N. Zhang, E. Du, Y. Cheng, C. Kang are with State Key Lab of Power Systems; Dept. of Electrical Engineering; Tsinghua University; Beijing, China, 100084.

D. S. Kirschen is with the Dept. of Electrical and Computer Engineering, University of Washington, Seattle, WA 98195.

## II. Literature Review

As mentioned above, a number of authors have developed techniques to learn security rules using decision trees. For instance, Sun et al. trained a DT offline and used the learned security rules for online dynamic security assessment with real-time PMU measurements. The results showed that DTs can identify key security indicators for large real-world power system [7]. Diao et al. also trained a DT offline for online voltage security assessment and updated periodically to improve real-time performance and robustness [9]. Other researchers use the DT rules for preventive and corrective control [17][18][19]. For example, Genc et al. utilize the DT to learn security regions and boundaries and provide guidelines for preventive and corrective controls, such as load shedding and generation re-dispatch [17]. Voumvoulakis et al. train two DTs for power system dynamic security assessment and corrective control, respectively [18]. DT rules have also been applied to power system cascading events [20] and emergency control [21] [22]. For instance, Bernabeu et al. developed a DT-based scheme to reduce the possibility of hidden failures and cascading events by adjusting the balance between security and dependability [20]. Senroy et al. developed DT rules using simulation data to identify pre-conditions when a power system should be split into islands [21]. Note that these applications to security assessment, as well as preventive corrective and emergency control were conducted in a context of a low penetration of renewable generation.

Recently, some authors have investigated how security rules could be embedded in economic dispatch model to handle the increasing penetration of variable generation. For instance, Thams et al. presented a data-driven security-constrained optimal power flow framework. They first use the DT to learn rules for both small signal stability and steady-state security, and then embedded these rules as constraints in a DCOPF to minimize re-dispatching actions [11]. Halilbasic et al. extend this method to embed the security rules in an ACOPF. They proposed a second-order cone relaxation to transform the security-constrained problem into a solvable convex mixed-integer OPF formulation [12]. Cremer et al. used ensemble decision tree (AdaBoost) to learn the power system security boundary. They relied on generalized disjunctive programming to extract DT rules and then embedded the rules in a DCOPF to balance cost and risk [13].

## III. Data-Driven Framework

Security rules derived from a DT must satisfy three fundamental requirements:
1. Highly accurate: the rules must reflect as accurately as possible actual system operating conditions. To achieve this, the DT must have a high learning capacity and they should be easily grouped into ensembles when necessary.
2. Understandable: power system operators must be able to obtain an intuitive understanding of the justification for each rule. To achieve this goal, the DT should be sparse, have low depth, and be able to learn the correlations among power system features.
3. Embeddable in optimization: some security or stability constraints do not have an analytical form that can be embedded in optimization problems. Therefore, the DT should be able to learn complex security boundaries and extract linear rules, which can be introduced as constraints in these optimization problems. Embeddable rules should involve a small number of integer variables and nonzero parameters. In other words, there should be as few as possible leaves in a single sparse DT.

To satisfy these requirements, this paper uses the three-stage, data-driven framework for rules extraction based on a sparse oblique decision tree illustrated on Fig. 1.

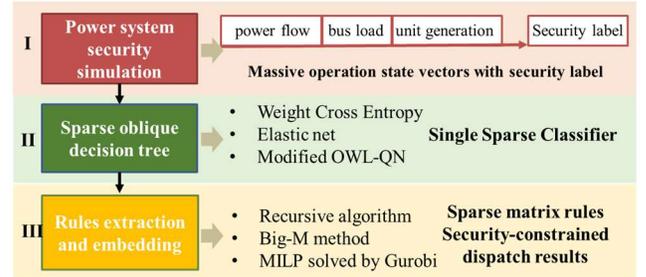

Fig. 1. Framework for power system security rules extraction and embedding.

### A. Stage I---Generation of a large dataset of power system operating states through simulation

Since actual datasets of insecure power system operating states are rare, simulation results are needed to generate sufficient data. Without loss of generality, this paper focuses on N-k security analysis in power system with high renewable penetration. First, wind and PV generations are simulated using stochastic differential equations considering spatial-temporal correlations [6]. Next, an economic dispatch is run to determine the operating state. Line outage distribution factors (LODF) are then applied to evaluate the N-k security of this state [23]. This process is repeated to generate a dataset $P$ of power system states with security labels $y$:

$$P = \{(p_1, y_1), (p_2, y_2), \dots, (p_N, y_N)\} \quad (1)$$

An operating state vector $p$ includes the power flow $l$, unit generation $g$, renewable energy generation $r$, and load $d$ features:

$$p = (g^T, r^T, l^T, d^T, 1)^T \quad (2)$$

where the last feature is always "1" to learn offset. The labels $y \subset S$, where $S$ is the set of security labels, which in this paper is $S = \{0, 1\}$, where 0 denotes an insecure state and 1 denotes a secure state. It should be noted that the features of the operating state are not unique and are chosen depending on the type of security problem considered.

*B. Stage II---Training of a sparse oblique decision tree*

The dataset of security-labeled operating states is high-dimensional and correlated. To learn sparse and linear security rules that satisfy the three requirements highlighted above, we develop an algorithm to build sparse weighted oblique decision trees (SWODT) that can be used as a classifier for power system security assessments. An oblique decision tree has an improved learning capacity because it splits at each node based on a linear combination of values of several features rather than on the value of a single feature. Sparsity in the parameters of this linear split is achieved by minimizing an impurity index and elastic net regularization.

*C. Stage III—Extraction of the rules and embedding in economic dispatch*

Sparse rules are extracted from the SWODT using a recursive algorithm. Each leaf of the decision tree with a secure label is extracted as a sparse matrix that reflects the direction of the split at each node. This matrix is then embedded as security constraints in the economic dispatch using the Big-M method. This security constrained economic dispatch problem is a MILP, which can be solved using a standard software package.

## IV. SPARSE WEIGHTED OBLIQUE DECISION TREE

*A. Weighted oblique decision tree (WODT)*

Oblique decision trees have an expanded learning capacity because they perform at each node a binary test using a linear split function $\theta^T p$ rather than testing the value of a single feature at each node. If $\theta^T p < 0$, the power system state $p$ belongs to the left child node; otherwise, to the right child node. In the basic form, at each node the vector of parameters $\theta$ is trained to minimize an impurity index. However, this approach does not converge well because the indicator function $I[\theta^T p < 0]$ is discontinuous. To this end, Yang et al. developed weighted oblique decision trees (WODT) with weighted information entropy as an impurity index, which uses the smooth "soft split" sigmoid function $\sigma(\theta^T p)$ rather than the "hard split" indicator function $I[\theta^T p < 0]$ [24]. We propose to further combine the weighted information entropy with the elastic net [25] as the objective function to derive split parameters of the current node. While the weighted information entropy makes the impurity index smooth and easy to optimize, the elastic net sets the parameters of unimportant and irrelevant features to zero while keeping the correlations among the important features. Given that the objective function is smooth in each orthant, we modify the OWL-QN method to solve the non-smooth optimization problem and recursively obtain the whole sparse weighted oblique decision tree. The weights of sample $p$ directed to the left $w^L$ and right $w^R$ child nodes are calculated as follows:

$$w^L = \sigma(-\theta^T p) \\ w^R = \sigma(\theta^T p) \quad (3)$$

To obtain split parameters at each node of the WODT, the objective function is:

$$Min_\theta \ \tilde{E}(\theta) = W_L(\theta)H_L(\theta) + W_R(\theta)H_R(\theta) \quad (4)$$

where $W_L(\theta)$ and $W_R(\theta)$ are the sum of the weights of the left and right child nodes:

$$W_L(\theta) = \sum w_i^L \\ W_R(\theta) = \sum w_i^R \quad (5)$$

$H_L(\theta)$ and $H_R(\theta)$ are the left and right information entropies:

$$H_L(\theta) = -\sum_{k=1}^{|S|} \frac{W_L^k(\theta)}{W_L(\theta)} \log_2 \frac{W_L^k(\theta)}{W_L(\theta)} \\ H_R(\theta) = -\sum_{k=1}^{|S|} \frac{W_R^k(\theta)}{W_R(\theta)} \log_2 \frac{W_R^k(\theta)}{W_R(\theta)} \quad (6)$$

where $W_L^k(\theta)$ and $W_R^k(\theta)$ are the sum of the weights of the left and right child nodes with label $k$; $|S|$ is the number of security categories.

The unconstrained smooth optimization (4) can be solved using the L-BFGS algorithm. The WODT outperforms state-of-art oblique decision tree algorithms such as OC [26] and CART-LC [14] [24]. However, some features of WODT limit its application in the extraction of power system security rules:

*Limitation 1:* Because the split parameters $\theta$ are dense and the power system data are high-dimensional, the security rules are impossible to understand, and hard to be embedded in power system optimization.

*Limitation 2:* While the objective function is smooth, it is not convex, which affects the convergence of the L-BFGS algorithm.

*Limitation 3:* $\theta = 0$ is always a stationary point of the optimization problem. When the initialization is poor, the quasi-newton algorithm converges to this meaningless point, which prevents the algorithm from further splitting the current node. This problem is more significant when $\theta$ is sparse.

*B. Sparse weighted oblique decision tree (SWODT)*

To extract understandable and embeddable security rules from the high-dimensional and highly correlated power system dataset, we propose a sparse weighted oblique decision tree. At each node of a SWODT, the objective function is:

$$E(\theta) = \frac{1}{N}\{W_L(\theta)H_L(\theta) + W_R(\theta)H_R(\theta)\} + \lambda_1|\theta| + \lambda_2\|\theta\|^2 \quad (7)$$

The first term of this function is the impurity index weighted information entropy divided by the number of samples $N$ at the current node. The second term provides sparsity using the Lasso regularization function. The third term is the Ridge regularization function. The sum of the second and third parts is the Elastic Net [25]. Although the Lasso term alone can improve sparsity, the Ridge term is required for two main reasons:

(1) Power system features such as line flows are highly correlated. Lasso tends to ignore correlations and select only the most important feature. Lasso also does not work when the number of features is greater than the number of samples. Adding Ridge can help the Elastic Net capture correlations among features and make the Lasso working with a large number of features while keeping the parameters sparse [25].

(2) The weighted information entropy function is nonconvex, while Ridge is convex everywhere in domain. Improving the convexity helps the quasi-newton algorithm converge faster. If we denote the Hessian matrix of the first term of $E(\theta)$ by $H$, the overall Hessian matrix on the current orthant is then $H + 2\lambda_2 I$ and all eigenvalues of $H$ are increased by $2\lambda_2$, which makes the Hessian matrix of $E(\theta)$ more likely to be positive definite.

According to Eq.(6), the ranges of $H_L(\theta)$ and $H_R(\theta)$ are:

$$0 \leq H_L(\theta) \leq \log_2(|S|)$$
$$0 \leq H_R(\theta) \leq \log_2(|S|) \quad (8)$$

Therefore, the range of weighted information entropy $\tilde{E}(\theta)$ is:

$$0 \leq W_L(\theta)H_L(\theta) + W_R(\theta)H_R(\theta) \leq N\log_2(|S|)$$
$$W_L(\theta) + W_R(\theta) = N \quad (9)$$

Eq. (9) shows that the range of $\tilde{E}(\theta)$ depends on the number of samples. Because the number of samples is different at each node of SWODT, we would have to choose a suitable regularization coefficient at each one, which is intractable. Instead, we divide the first part of $E(\theta)$ by $N$ to reduce this range to $[0, \log_2(|S|)]$, which makes it independent of the number of samples at each node.

The objective function (7) is non-smooth due to the Lasso term and thus cannot be solved using the L-BFGS method. However, the function is smooth on each orthant. Algorithm 1 shows how the Orthant-Wise Limited-memory Quasi-Newton (OWL-QN) method [27] has been modified to solve this non-smooth optimization problem by taking advantage of this feature.

Let: $f(\theta) = \frac{1}{N}\{W_L(\theta)H_L(\theta) + W_R(\theta)H_R(\theta)\} + \lambda_2\|\theta\|^2 \quad (10)$

Then: $E(\theta) = f(\theta) + \lambda_1|\theta| \quad (11)$

OWL-QN limits every optimization step to the current orthant. Then, the Hessian matrix of $E(\theta)$ on the current orthant is determined only by the smooth part $f(\theta)$. Therefore, the traditional L-BFGS algorithm can be used to

---

**Algorithm 1** modified OWL-QN

**Input:** initialized $\theta \leftarrow \theta_0$, $S \leftarrow \{\}$, $Y \leftarrow \{\}$, $\varepsilon$

1: **for** $k = 0$ to MaxIters **do**
2:    Calculate pseudo derivative $v_k = -\lozenge E(\theta_k)$ with Eq. (12)
3:    **if** $(Y)^T S > \varepsilon$ **then**
4:      Update Hessian matrix $\mathbf{H}_k$ and calculate descent direction $d_k = \mathbf{H}_k v_k$ with $S$, $Y$ using L-BFGS algorithm
5:      Calculate constrained descent direction $q_k = \pi(d_k; v_k)$
6:    **else**
7:      Calculate descent direction $q_k = v_k$
8:    **end if**
9:    Find a suitable step size $\alpha = \beta^m$ and parameter $\gamma$ with constrained backtracking line search to satisfy:
$$E(\theta_{k+1}) \leq E(\theta_k) - \gamma v_k^\top(\theta_{k+1} - \theta_k)$$
$$\theta_{k+1} = \pi(\theta_k + \alpha q_k; \xi_k)$$
10:   **if** termination condition satisfied **then**
11:     Stop and return $\theta_{k+1}$
12:   **end if**
13:   Update $S$ with $s_k = \theta_{k+1} - \theta_k$
14:   Update $Y$ with $y_k = \nabla f(\theta_{k+1}) - \nabla f(\theta_k)$
15: **end for**

**Output:** split parameters $\theta$

---

update the Hessian matrix at the current step and calculate descent direction $d_k$ (Step 4). Next, the constrained descent direction $q_k$ is calculated by constraining $d_k$ with pseudo derivative $\lozenge E(\theta)$ (Step 5). $\lozenge E(\theta)$ is used to replace the original derivative considering the non-smooth feature of Lasso (Step 2):

$$\lozenge_i E(\theta) = \begin{cases} \nabla_i f(\theta) + \lambda_1, & \text{if } \theta^i > 0 \\ \nabla_i f(\theta) - \lambda_1, & \text{if } \theta^i < 0 \\ \nabla_i f(\theta) + \lambda_1, & \text{if } \theta^i = 0, \nabla_i f(\theta) + \lambda_1 < 0 \\ \nabla_i f(\theta) - \lambda_1, & \text{if } \theta^i = 0, \nabla_i f(\theta) - \lambda_1 > 0 \\ 0, & \text{otherwise} \end{cases} \quad (12)$$

where the derivative of $f(\theta)$ is:

$$\nabla f(\theta) = \frac{1}{N}\sum_{i=1}^{N}\left\{\sigma(\theta^T p_i)[1 - \sigma(\theta^T p_i)]p_i \log_2 \frac{W_R W_L^{y_i}}{W_L W_R^{y_i}}\right\} + 2\lambda_2 \theta \quad (13)$$

Next, a backtracking line search is conducted to find the next point along the descent direction $q_k$ (Step 9). If the next point crosses the current orthant in the line search, it will be projected back to the current orthant ($\theta_k + \alpha q_k$ to $\theta_{k+1}$ in Fig. 2). The projecting function $\pi$ is as follows:

$$\pi_i(\theta_k + \alpha q_k; \xi_k) = \begin{cases} \theta_k^i & \text{if } \delta(\theta_k^i) = \delta(\xi_k^i) \\ 0 & \text{otherwise} \end{cases} \quad (14)$$

where $\delta(\cdot)$ is sign function, $\xi_k$ is the current orthant.

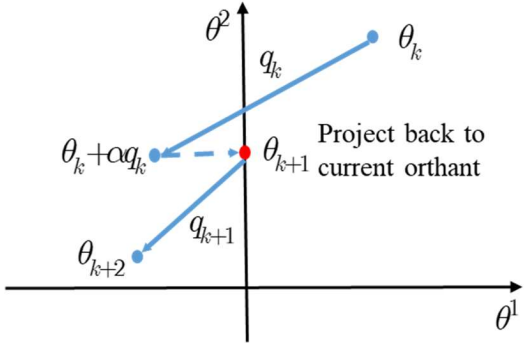

Fig. 2. Illustration of the OWL-QN algorithm

The current orthant $\xi_k$ is determined by the sign of the current point and pseudo derivative $\Diamond E(\theta_k)$:

$$\xi_k^i = \begin{cases} \delta(\theta_k^i) & \text{if } \theta_k^i \neq 0 \\ \delta(-\Diamond_i E(\theta_k)) & \text{if } \theta_k^i = 0 \end{cases} \quad (15)$$

The backtracking line search in OWL-QN guarantees a sufficient descent at each step but may fail to satisfy $(Y)^T S > 0$, which is necessary to make the Hessian matrix positive definite in the L-BFGS method. The non-positive definite Hessian matrix may cause $q_k = 0$ when constrained by the pseudo derivative. Therefore, in this paper, if $(Y)^T S < \varepsilon$ (a very small positive number), the pseudo derivative is used as the descent direction (Steps 3 and 7). This means that we adopt the first-order descent method when the second-order descent method is not suitable at the current point. Adding the strong convexity of the Ridge term reduces the risk of $(Y)^T S < 0$ after the backtracking line search and accelerates convergence.

Algorithm 2 shows how to construct the SWODT by obtaining the split parameters at each node using the modified OWL-QN recursively. First, if the current node is not identified as a leaf, then the recursion continues (step 2-7). Since a UDT provides acceptable performance as well as sparsity, we use it to initialize the split parameters to accelerate the convergence of the proposed quasi-newton method (Step 8). The split parameters are calculated by solving the unconstraint optimization problem (7) using the modified OWL-QN (Step 9). If $\lambda_1$ is too large, the SWODT may still converge to the meaningless stationary point $\theta = 0$. In such cases, we replace the SWODT by the UDT at that node to improve robustness and ensure that the performance is no worse than with the UDT (Step 10). Then, we obtain the training data $L, R$ for the left and right child nodes with hard split ($I[\theta^T p < 0]$). The process continues recursively for the left and right child nodes (Steps 11-13) until reaching a leaf node.

**Algorithm 2** SparseSubtree($P, D, d, J$) of the Sparse Weighted Oblique Decision Tree

**Input:** Training data $P$, maximum depth $D$, depth of current node $d$, minimum number of sample to split $J$
1: Create a node $Tr$ based on training data $P$
2: **if** $len(P) \leq J$ or $d > D$ **then**
3: Node $Tr$ is a leaf node labeled as majority class $k$ in $P$
4: **end if**
5: **if** all samples in $P$ belongs to same class $k$ **then**
6: Node $Tr$ is a leaf node labeled with class $k$
7: **end if**
8: Initialize the split parameters $\theta_0$ with univariate decision tree
9: Calculate the split parameters $\theta$ by solving the unconstraint optimization problem (7) using Algorithm 1 with initial point $\theta_0$
10: **if** $\theta = 0$ **then** $\theta = \theta_0$ **end if**
11: Obtain training data $L, R$ for left and right child nodes, respectively
12: The left child node $Tr$: SparseSubtree($L, D, d + 1, J$)
13: The right child node $Tr$: SparseSubtree($R, D, d + 1, J$)

**Output:** A sparse oblique decision tree with root node $Tr$

## V. RULES EXTRACTION AND EMBEDDING

### A. Rules extraction

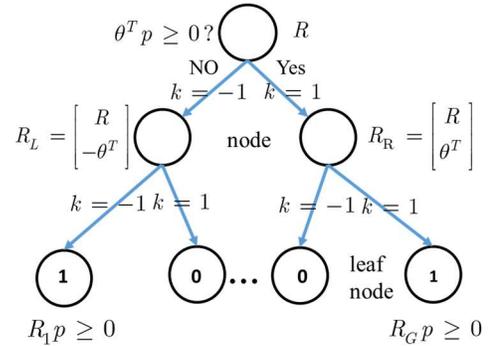

Fig. 3. Illustration of the rule extraction algorithm

Algorithm 3 and Fig. 3 show how to extract the sparse rules matrix recursively from the SWODT. If the current node is identified as a leaf node with secure label "1", then the security rules of the leaf are extracted as a sparse matrix $R$ (Steps 1-3). Otherwise, for each node, the current split considering direction $k\theta^T$ is added to the last row of rules matrix $R$ to represent the rules of left and right child nodes respectively (Step 5). $k = -1$ and $k = 1$ denote respectively the left split ($\theta^T p < 0$) and right split ($\theta^T p \geq 0$), as shown in Fig. 3. The process continues recursively for the left and right subtrees until reaching a secure leaf (Steps 7-11). If $R_i$ denotes the sparse rules matrix of the $i^{\text{th}}$ secure leaf of the SWOT, the corresponding rules can be expressed as:

$$R_i p \geq 0 \quad (16)$$

| **Algorithm 3** RuleFromSubtree(*Tr, R*) of SWODT |
|---|
| **Input:** SWODT node *Tr* |
| 1:     **if** *Tr* is a leaf node **and** the class label is 1 **then** |
| 2:        **return** *R* |
| 3:     **end if** |
| 4:     **for** $k$ **in** $\{-1,1\}$ |
| 5:        Concatenate vector $k\theta^T$ to the row of matrix $R$ |
| 6:        **if** $k = -1$ **then** |
| 7:           Extract rules from left subtree *Tr_left*: RuleFromSubtree(*Tr_left, R*) |
| 8:        **end if** |
| 9:        **if** $k = 1$ **then** |
| 10:       Extract rules from right subtree *Tr_right*: RuleFromSubtree(*Tr_right, R*) |
| 11:     **end if** |
| 12:    **end for** |
| **Output:** Sparse rules matrix *R* |

### B. Embedding security rules in economic dispatch

Given Eq.(16), if the SWODT has $G$ secure leaves, then the sparse security rules of the whole tree can be formulated as security constraints using Big-M method:

$$\begin{aligned} R_1 p &\geq -M(1-I_1) \\ R_2 p &\geq -M(1-I_2) \\ &\cdots \\ R_G p &\geq -M(1-I_G) \\ \sum_{i=1}^{G} I_i &= 1 \end{aligned} \quad (17)$$

where $I_i$ is the binary variable representing $i^{\text{th}}$ leaf of the SWODT and *M* is a sufficiently large positive number. In Eq. (17), the last equation indicates that only one rule needs to be satisfied to obtain a secure power system state. However, this equation is not actually needed because any sample belongs to one and only one leaf. Embedding Eq. (17) in an economic dispatch yields a secure system operating state. Since these equations are linear, this security-constrained economic dispatch can be solved using a standard MILP solver if DCOPF is used. This solution does not require too much computing time because Eq. (16) is sparse.

## VI. CASE STUDIES

### A. Description of the case data

Three power system test systems (IEEE-30, RTS-GMLC [28], NREL-118 [29]) with high renewable penetrations are used in this paper to validate the proposed method. Table I gives the structure and generation capacity of these three cases. The IEEE-30 case is modified from the standard IEEE-30 case. A 50 MW PV farm and a 50 MW wind farm were added at each one of buses #10, #24, and #28. The capacity of renewable energy is thus 300 MW, which is 47% of the total generation capacity. The RTS-GMLC case was created by NREL to replace RTS-96 for high renewable penetrated power system reliability assessment. The hydro, wind, and PV generation capacities are 1000 MW, 2508 MW, and 2716 MW, respectively, accounting for 43% of the total capacity. Day ahead forecasts and actual time-series data of hydro, wind, and PV production, as well as load based on actual power systems are provided. The NREL-118 dataset was also introduced by NREL to add a high renewable penetration to the original IEEE-118 case. The hydro, wind, and PV capacities are 10157 MW, 1078 MW, and 3446 MW, respectively, accounting for 36% of the total generation capacity.

8000 power flow simulations were performed on each dataset to obtain N-1 security-labeled samples to train decision trees. These three datasets are used to compare the original weighted oblique decision tree (WDOT), the proposed sparse weighted oblique decision tree with Lasso only as regularization and solved by modified OWL-QN (SWODTL), the proposed sparse weighted oblique decision tree (SWODT). Metrics of accuracy and sparsity, as well as number of leaves, and computing times were calculated on the three datasets using 5-fold cross validation.

TABLE I: Description of the three test cases

| Year | IEEE-30 | RTS-GMLC | NREL-118 |
|---|---|---|---|
| Hydro capacity (MW) | 0 | 1000 | 10157 |
| Thermal capacity (MW) | 335 | 8326 | 25837 |
| Wind capacity (MW) | 150 | 2508 | 1078 |
| PV capacity (MW) | 150 | 2716 | 3446 |
| Renewable capacity penetration | 47% | 43% | 36% |
| Peak load (MW) | 265 | 8192 | 18063 |
| Number of buses/branches | 30/43 | 73/120 | 118/186 |
| Number of total/renewable generators | 12/6 | 158/60 | 160/92 |

### B. Accuracy and training time

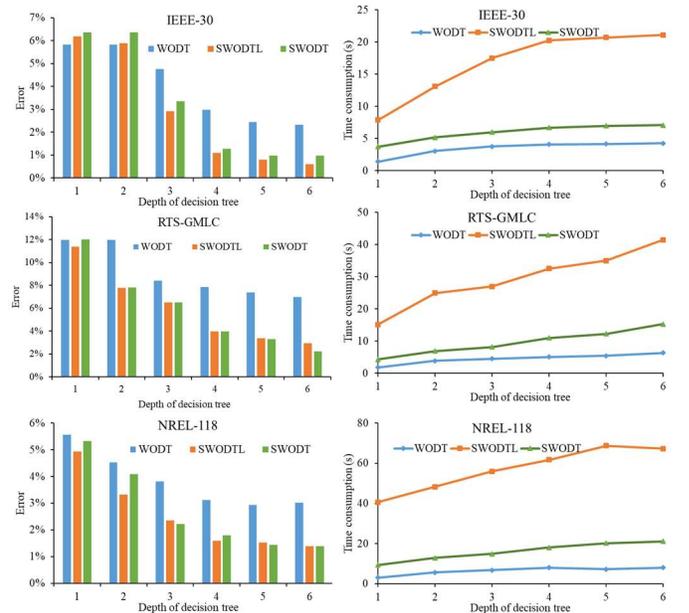

Fig. 4. Accuracy and training time of the three oblique decision trees for the three datasets

Fig. 4 shows the accuracy and training time of the three oblique DTs (WODT, SWODTL, and SWODT) against the depth of the DT on three datasets. The error rate is defined as the average ratio of the numbers of insecure and total states in 5-fold validation dataset. When the depth of the DT is low (depth 1 and 2 in IEEE-30 and NREL-118, depth 1 in RTS-GMLC), the accuracy of the three methods is comparable, possibly due to underfitting. When the depth of DT increases, the accuracy of SWODT and SWODTL is similar and both have much better performance (approximately 2%) than the original WODT (such as depth 3-6 in IEEE-30 and NREL-118, depth 2-6 in RTS-GMLC). This is because the Lasso and Ridge regularization avoid possible overfitting and reduce generalization errors by making the security rules simple when the DT is deep. These results suggest that oblique DTs have a strong capacity to learn security rules.

The average training time of the three methods increases approximately linearly with the depth of the trees. However, the training time required by the SWODTL is about 5 times larger than for the WODT and 3 times larger than for the SWODT because the strong non-convexity of SWODTL objective function makes the modified OWL-QN algorithm converge slowly. The training times of the SWODT and WODT are close because the Ridge accelerates SWODT's convergence, although the modified OWL-QN limits the update on current orthant. These results show that SWODT has better performance than WODT while consuming much less training time than SWODTL.

*C. Sparsity and number of leaves*

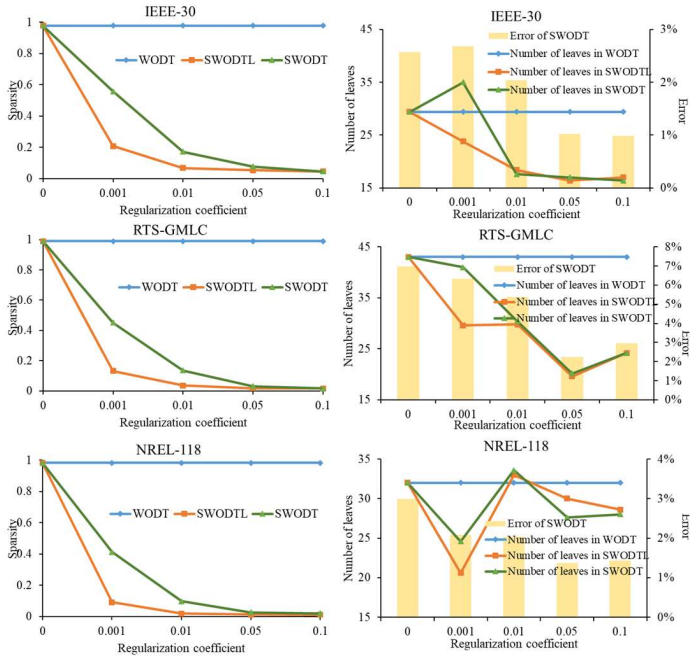

Fig. 5. Sparsity and number of leaves of the three oblique decision trees for the three datasets

Fig. 5 compares the sparsity and the number of leaves of the three oblique DTs (WODT, SWODTL, and SWODT) against the regularization coefficient on the three datasets. The regularization coefficient of SWODT is set as $\lambda_1 = \lambda_2$.

The depth is set as 6 for all DTs. Sparsity in this paper is defined as the proportion of non-zero parameters in the rule matrix. The sparsity of the WODT is approximately 1 for all three datasets, which means that its rules matrix is very dense. The security rules are then complex and impossible to understand. The sparsity of SWODTL and SWODT decreases as the regularization coefficient increases and approaches 0 for a sufficiently large regularization coefficient. The security rules derived from SWODTL and SWODT can therefore be made very simple by choosing a suitable regularization coefficient. The sparsity of the SWODTL decreases faster than that of the SWODT because the SWODTL ignores the correlations and only keeps the parameters of the most important feature. The WODT has an average of 29.4, 43, and 32 leaves on the IEEE-30, RTS-GMLC, and NREL-118 datasets, respectively, nearly twice as many as for the SWODT and SWODTL when these use a suitable regularization coefficient. A smaller number of leaves leads to fewer rules matrices and fewer integer variables that must be embedded in the security constrained economic dispatch. The small number of leaves and sparsity also make the SWODT much easier to be understood than WODT. The graphs on the right also show that adjusting the regularization coefficient also reduces the error rate. For these three datasets, the suitable coefficients are 0.05, 0.05 and 0.001 for the IEEE-30, RTS-GMLC, and NREL-118, respectively.

*D. Embedding Rules*

To validate the effectiveness of the extracted security rules, we embedded the rules as constraints in an economic dispatch for the three datasets, generated 2000 new operating states by randomly changing renewable energy generation and load. We then assessed whether these states met the operating constraints and calculated the ratio of the numbers of secure and total states. As a reference, an economic dispatch without security constraints results in 62.2%, 76.0% and 52.9% of secure states for the IEEE-30, RTS-GMLC, and NREL-118 datasets, respectively. After embedding the rules from the three oblique DTs, the percentage of secure states produced by the economic dispatch improved from 62.2% to over 95% for the IEEE-30 (WODT: 95.0%, SWODTL: 95.1%, SWODT: 95.9%), from 76.0% to over 80.0% for the RTS-GMLC (WODT: 80.0%, SWODTL: 80.3%, SWODT: 85.1%), and from 52.9% to over 74.0% for the NREL-118 (WODT: 74.0%, SWODTL: 85.4%, SWODT: 93.5%). These results indicate that the SWODT outperforms both WODT and SWODTL because the dense structure limits the generalization capability of WODT, and SWODTL ignores the correlations among the power system features. Fig. 6 summarizes these results and also shows that, thanks to its sparse rules matrices and smaller number of integer variables, the SWODT also outperforms WODT and SWODTL based on the solution time averaged over all cases and all three datasets. The time difference between the WODT and the SWODT is more significant in a larger power system (0.05s on the IEEE-30 against 0.1s on the NREL-118). The simulation platform is a workstation with an Intel i7 CPU@2.11 GHz, 32 GB of RAM. The MILP economic dispatch problem is solved by Gurobi 8.0.

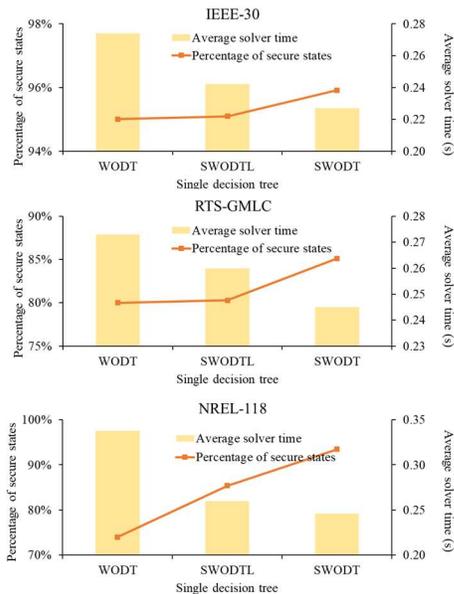

Fig. 6. Percentage of secure states and average solution time when embedding the security rules extracted from the WODT, SWODT, SWODTL in a security constrained economic dispatch for the IEEE-30, RTS-GMLC, and NREL-118 datasets

## VII. CONCLUSION

This paper proposes a technique based on sparse weighted oblique decision trees to learn security rules in power systems with a high penetration of variable renewable generation. The sparse weighted oblique decision tree defines accurate, understandable, and embeddable security rules that can be extracted as sparse matrices and then embedded in economic dispatch as linear security constraints. The effectiveness of the proposed method is validated on the IEEE-30, RTS-GMLC and NREL-118 datasets. The results show that the sparse weighted oblique decision tree outperforms the state-of-art weighted oblique decision tree while keeping the security rules sparse and simple. Embedding the sparse security rules in power system economic dispatch can significantly increase the percentage of insecure states and reduce the average solution time of the security-constrained economic dispatch.